\documentclass[twocolumn,showpacs,preprintnumbers,amsmath,amssymb]{revtex4}
\usepackage{graphicx}
\begin{document}
\preprint{IMAFF-RCA-02-08}
\title{Unified Model for Dark Energy}

\author{Pedro F. Gonz\'{a}lez-D\'{\i}az}
\affiliation{Centro de F\'{\i}sica ``Miguel A. Catal\'{a}n'', Instituto de
Matem\'{a}ticas y F\'{\i}sica Fundamental,\\ Consejo Superior de
Investigaciones Cient\'{\i}ficas, Serrano 121, 28006 Madrid (SPAIN).}
\date{\today}
\begin{abstract}
A new model for the universe filled with a generalized Chaplygin
fluid is considered which unitarily describes as a single vacuum
entity both a quintessence scalar field and a cosmological
constant, so unifying the notion of dark energy. While the
evolution of the universe filled with such a fluid does not
obviously contradict the present cosmic acceleration, the
introduced single dark-energy component, for equations of state
with characteristic parameter $\omega\geq -1$, behaves like an
usual quintessence fluid with constant equation of state at early
high densities, and like a pure cosmological constant at late
cosmological times.
\end{abstract}

\pacs{98.80.-k, 98.80.Es}

\maketitle

Rather surprising new evidence for dark energy from a ten-year
census of the gravitational lensing in distant quasars has been
quite recently added by Browne et al. [1] to the itself compelling
evidence originally obtained from distant supernova [2,3] and all
those ever favorable results later amassed in the past five years
[4]. Thus, it now appears as a rather solid conclusion that the
present universe is accelerating and that most of its energy is in
the form of a mysterious dark energy, which has long-range
anti-gravity properties. Dark energy is closely related to the
idea of a positive cosmological constant, or to the notion of the
so-called cosmic quintessence [5], a slowly-varying, tracked [6]
or not [7], scalar field that produces a negative pressure. But,
whatever its ultimate nature, dark energy has become one of the
main problems of all sciences [8].

On the other hand, dark energy has also been claimed [9] to be
just one of the two vacuum-energy components in a unified model
for {\it cosmic darkness} whose other component is dark matter.
Rather than fine-tuning any specific quintessence potential, this
model explains the present accelerating expansion of the universe
by means of the rather exotic equation of state that corresponds
to a generalized Chaplygin gas [10], acting in such a way that it
caused the gas to behave like pure dark matter at early times
(high density) and like dark energy at late times (small density).
The generalized Chaplygin gas is a substance which is
characterized by an equation of state [9-11]
\begin{equation}
p=-A\rho^{-\alpha} ,
\end{equation}
with $A$ a positive definite constant and $\alpha$ a parameter
which can take on positive or negative values. The original
Chaplygin gas was characterized by $\alpha=1$. It is an equation
of state with a form as given by Eq. (1) that has thus been
claimed to represent the stiff that simultaneously describes dark
matter and dark energy. In this letter we generalize the idea of
Chaplygin gas still one step further in such a way that, rather
than unitarily characterizing dark matter and dark energy as a
single entity, the resulting gas theory would now simultaneously
describe the physics of a quintessence scalar field and that of a
cosmological constant as the two opposite limiting situations that
result at early and late times, respectively, from a unique
unified dark energy physical picture.

The dynamical cosmological equations that represent a pure
quintessence field with constant state equation [7] $p=\omega\rho$
are known to be valid only in an $\omega$-interval,
$0\leq\omega<-1$, which cannot include the value $\omega=-1$ that
corresponds to a pure cosmological constant $\Lambda$ [12]. The
generalization of the Chaplygin gas that we are going to introduce
does neither require specifying any potential for the generalized
field and is based on suitably combining the state equations for a
slowly-varying quintessence scalar field with that for the
generalized Chaplygin gas given by Eq. (1), in such a way that the
resulting state equation would give rise to cosmological dynamical
equations that would represent both a quintessence-like field and
a cosmological constant in the above single unified way. We take
for that state equation the following simplest combination
\begin{equation}
p=\left(\omega-\frac{A}{\rho^{\alpha+1}}\right)\rho ,
\end{equation}
where $p$ and $\rho$ are respectively pressure and energy density
in a comoving reference frame, with $p<0$ and $\rho>0$,
$0\geq\omega >-1$ and $A$ is again a constant, but one which
should not be necessarily positive definite, provided $p\leq 0$.
This equation of state is interesting as it can be connected with
both cosmology through parameter $\omega$ and string theory
through its Chaplygin-like gas component [9]. Inserting Eq. (2)
into the general cosmic conservation law for energy [13]
\begin{equation}
d\rho=-3\left(\rho+p\right)\frac{da}{a} ,
\end{equation}
we obtain for the density of the dark energy Chaplygin gas a
nonlinear expression in terms of the scale factor $a$. It is:
\begin{equation}
\rho(a)=\left\{\frac{1}{1+\omega}\left[A+\frac{B}{a^{3(1+
\alpha)(1+\omega)}}\right]\right\}^{1/(1+\alpha)} ,
\end{equation}
with $B$ an integration constant and $a\equiv a(t)$ the cosmic
scale factor which is normalized to unity today $a=(1+z)^{-1}$,
when expressed in terms of the redshif $z$. A differential feature
of the present model in relation with the unified dark matter
model is that, although the density never drop below the value
$A^{1/(1+\alpha)}$, no matter how much you expand the gas, it get
a value $\rho\propto a^{-3(1+\omega)}$ when sufficiently
compressed, so that the density can never reach the value
$\rho\propto a^{-3}$ if $\omega <0$.

Let us now introduce the re-definitions for the current
generalized dark-energy field $\phi$ cosmological parameter
\[\Omega_{\phi}^0=\frac{B}{A+B} .\]
and energy density
\[\rho_0=\left[\frac{A+B}{1+\omega}\right]^{1/(1+\alpha)} .\]
In terms of these quantities Eq. (4) can be re-written in a form
which is similar to that has been used in unified dark matter
models [14]
\begin{equation}
\rho(a)=\rho_0\left[1-\Omega_{\phi}^0
+\frac{\Omega_{\phi}^0}{a^{3(1+\omega)(1+\alpha)}}\right]^{1/(1+\alpha)}
.
\end{equation}
In order to interpret this equation, let us consider the
corresponding expression for the energy density in the case that
we have a quintessence slowly-varying scalar field and a
cosmological constant which are separatedly defined in the flat
cosmological case [15], i.e.
\begin{equation}
\rho(a)=\rho_0\left[\left(1-\Omega_{\Lambda}\right)a^{-3(1+\omega_0)}
+\Omega_{\Lambda}\right] .
\end{equation}
We first note that, although Eq. (5) involves a single substance
and Eq. (6) involves two, the energy densities given by these two
equations become both the density $\rho_0$ for the current value
($z=0$) of the scale factor $a=1/(1+z)=1$. Making then the
identifications
\[\Omega_{\phi}^0=\left(1-\Omega_{\Lambda}\right)^{1+\alpha},\;\;\;
\omega=\omega_0 ,\] we see that Eqs. (5) and (6) again become the
same and are both given by
\[\rho=\rho_0\Omega_{\phi}^{0}a^{-3(1+\omega)} \]
in the limit $a\rightarrow 0$. It follows that $\Omega_{\phi}$
behaves like though it were an effective quintessence density in
the considered model. Also for the flat case, once again Eqs. (5)
and (6) become the same when $\alpha=0$ and are both given by
\[\rho=\rho_0\left[\Omega_{\phi}^0+\Omega_{\Lambda}a^{-3(1+\omega)}\right],
\]
which corresponds to the standard combined
"$\Lambda$-Quintessence" model which involves two substances.
Finally, Eqs. (5) and (6) become also the same in the limit
$a\rightarrow\infty$, where both read
\[\rho=\rho_0\Omega_{\Lambda} ,\]
in case that the identification
$\Omega_{\Lambda}=\left(1-\Omega_{\phi}^0\right)^{1/(1+\alpha)}$
be also introduced. This latter case would clearly correspond to
that for a pure cosmological constant.

The interpretation of our present picture for dark energy which
describes quintessence and cosmological constant in a unified,
single fashion can be further strengthened if we also introduce
definitions for the effective parameter entering the equation of
state and the effective speed of sound, which in the present case
become
\begin{equation}
\omega^{{\rm eff}}=\frac{p}{\rho}=\omega-\frac{(1+\omega)A}{A
+\frac{B}{a^{3(1+\omega)(1+\alpha)}}}
\end{equation}
\begin{equation}
c_s^{{\rm eff}2}=\frac{\partial
p}{\partial\rho}=\omega+\frac{(1+\omega)\alpha A}{A
+\frac{B}{a^{3(1+\omega)(1+\alpha)}}}
\end{equation}
We notice that both $\omega^{{\rm eff}}$ and $c_s^{{\rm eff}2}$
reduce to $\omega$ (the typical pure quintessential case) when
$a\rightarrow 0$, and to $\omega^{{\rm eff}}=-1$, $c_s^{{\rm
eff}2}=\alpha+\omega(1+\alpha)$ (the cosmological constant case)
whenever $a\rightarrow\infty$. This closes up the proof that our
model consistently describes both quintessence and cosmological
constant as given limiting concepts from a more general unified
dark energy description involving one single substance, and can
therefore be looked at as being a promising concept for cosmology.

However, a wide class of the so-called unified dark energy models,
which are similarly constructed using a less generalized Chaplygin
equation of state, has been recently excluded [14] because such
models give rise to oscillations or exponential browup in the
matter power spectrum that are inconsistent with observations. One
would in fact expect these unwanted phenomena to occur as well in
our present unified dark energy model. Following Sandvik et al.
[14], the evolution of density perturbations $\delta_k$ with wave
vector $k$ can be written in the form [16]:
\begin{eqnarray}
&&\delta_k ''+\left[2+\xi-3\left(2\omega^{{\rm eff}}-c_s^{{\rm
eff}2}\right)\right]\delta_k '\nonumber\\
&&-\left[\frac{3}{2}\left(1-6c_s^{{\rm eff}2}+8\omega^{{\rm eff}}
-3\omega^{{\rm eff}2}\right)\right.\nonumber\\
&&\left.-\left(\frac{kc_s^{{\rm
eff}}}{aH}\right)^2\right]\delta_k=0 ,
\end{eqnarray}
where $H$ is the Hubble parameter,
\[\xi=-\frac{2}{3}\left[1+\left(\frac{1}{\Omega_{\phi}^0}-
1\right)a^{3(1+\alpha)}\right]^{-1} ,\] and the prime denotes
differentiation with respect to the timelike independent variable
$\ln a$. By inspecting Eq. (9) it can in fact be inferred that
whenever the effective speed of sound is nonzero, the unwanted
oscillatory ($c_s^{{\rm eff}}>0$) and exponential blowup
($c_s^{{\rm eff}}<0$) behaviours of fluctuations with wavelengths
below the Jeans scale $\sqrt{\pi|c_s^{{\rm eff}}|^2/G\rho}$ would
always take place. Whereas this situation is unavoidable for the
wide class of unified dark matter models pointed out by Sandvik et
al. [14], here one always has the possibility to set $c_s^{{\rm
eff}}=0$ at all times, while keeping a nonzero value for the bare
speed of sound,
\[c_s^2=\frac{A\alpha}{A+\frac{B}{a^{3(1+\omega)(1+\alpha)}}},\]
because the constant $A$ may in our case take also on negative
values while the right-hand-side of Eq. (8) has two terms to play
with. Thus, restricting to the simplest case where we choose
$\alpha=-1$, and hence $\omega^{{\rm eff}}=0$, the fluctuations
appear to be free from the unwanted oscillatory and exponential
blowup behaviours. In this case, one has to take negative values
for the cosmological constant $A$ [a case which can have very
positive implications (see later on)], so that Eq. (9) reduces to:
\begin{equation}
\delta ''+\left(2-\frac{3}{2}\Omega_{\phi}^0\right)\delta
'-\frac{3}{2}\delta=0 .
\end{equation}
The solutions to Eq. (10) are
\[\delta_{\pm}=a^{\frac{3}{4}\Omega_{\phi}^0
-1\pm\sqrt{1+\frac{9}{16}\left(\Omega_{\phi}^0\right)^2-
\frac{3}{2}\left(\Omega_{\phi}^0-1\right)}} \] at all times. We
have thus shown that the unstability problem and the presence of
unobserved oscillations affecting most of the unified dark matter
models can both be avoided in the kind of models considered here
for unified dark energy.

All the above kinds of models are however unrealistic, unless as
asymptotic approximations, because they describe an asymptotic
universe devoid of observable matter. One therefore ought to
include some kind of observable matter and/or radiation contents
in order to make the models more realistic. This will be later
implemented by considering a FRW scenario where, besides the
unified dark energy field, a radiation field will also be
included. We shall classically describe such a radiation field as
a massless scalar field which is conformally coupled to gravity
[15,17]. Before doing that, we shall first consider the simplest
case of a spatially flat universe filled only with the single
generalized Chaplygin gas discussed in this letter that satisfies
the state equation given by Eq. (2). The Friedmann equation for
this problem is
\begin{equation}
\left(\frac{\dot{a}}{a}\right)^{2(1+\alpha)}=\frac{1}{1+\omega}\left[A+\frac{B}{a^{3(1+
\alpha)(1+\omega)}}\right] ,
\end{equation}
where the overhead dot means time derivative. As pointed out
above, the solution to Eq. (11) for the general case in which
$\alpha\neq 0$ can only match the solution to the Friedmann
equation for the simple case that corresponds to a quintessence
field with constant equation of state $p=\omega\rho$ plus a
cosmological constant $\Lambda=3H^2\lambda$, that is for e.g. an
allowed value of the initial, constant parameter $\omega$ whose
following evolution be able to reproduce the current cosmic
acceleration, $\omega=-2/3$,
\begin{equation}
a(t)=\frac{8\pi
G}{3\sigma\lambda}\sinh^2\left(\frac{1}{2}\sqrt{3\lambda}t\right)
,
\end{equation}
just in the limiting cases of small $a$ as $\alpha\rightarrow 0$,
with $A=\lambda$ and $B=8\pi G/(3\sigma)$, in which $\sigma$ is
another integration constant. Note that for negative cosmological
constant the above solution changes into
\begin{equation}
a(t)=\frac{8\pi
G}{3\sigma|\lambda|}\sin^2\left(\frac{1}{2}\sqrt{3|\lambda|}t\right)
.
\end{equation}
For the small values of $|\lambda|$ and $a$ for which this
oscillating solution is valid, it would be expected that the
universe will show [18] no future event horizon, a property which
has been claimed [19] to be of decisive importance for allowing a
consistent mathematical formulation of fundamental string and M
theories.

Applying the generalization considered in this letter to the
original Chaplygin gas ($\alpha=1$), one obtains from Eq. (11) for
the flat case in the simple situation that corresponds to the
initial equation of state parameter $\omega=-2/3$ (which is
compatible with the present cosmic accelerating expansion for
moderate values of $A/(A+B)$ and the most recent observational
constraints [20]),
\begin{equation}
\dot{a}^4 =3\left(Aa^4+Ba^2\right) .
\end{equation}
A closed-form general analytical solution to Eq. (14) can be
obtained for the scale factor $a$ in parametric form and is given
by
\begin{equation}
t=(3A)^{-1/4}\left\{\ln\left[\frac{1+F(a)}{1-F(a)}\right]^{1/2}
-\arctan\left[F(a)\right]\right\}+\zeta ,
\end{equation}
where $\zeta$ is an adjustable integration constant and
\[F(a)=\left(1+\frac{B}{Aa^2}\right)^{1/4}
=\left(1+\frac{\Omega_{\phi}^0}{\Omega_{\Lambda}^0
a^2}\right)^{1/4} .
\]
Choosing
\[\zeta=\frac{(1-i)\pi}{2(3A)^{1/4}}\]
for the integration constant so that $a\rightarrow 0$ as
$t\rightarrow 0$, we finally get
\begin{equation}
t=(3A)^{-1/4}\left\{\ln\left[\frac{F(a)+1}{F(a)-1}\right]^{1/2}
+\frac{\pi}{2} -\arctan\left[F(a)\right]\right\}
\end{equation}
It follows from this solution that as the single gas is evolving
from a pure quintessence field into a cosmological term, the scale
factor $a$ will increses from zero, at $t=0$, to infinity, as
$t\rightarrow\infty$, such as one should expect in the unified
picture for dark energy.

Let us next include a term in the Friedmann equation that accounts
for the existence of an observable radiation field. As we
mentioned before, we represent this field by means of a massless
scalar field which is conformally coupled to gravity [17] so that
one has to modify [21] the Friedmann equation (11) to read:
\begin{equation}
\left[\left(\frac{\dot{a}}{a}\right)^2-\frac{M^2}{a^4}\right]^{1+\alpha}=
\frac{1}{1+\omega}\left[A+\frac{B}{a^{3(1+
\alpha)(1+\omega)}}\right] ,
\end{equation}
where $M^2$ is an integration constant. Although we have been
unable to find any analytical solution to Eq. (17) in closed-form,
it is still possible to readily deduce that in the limit of very
small values of the scale factor we have the approximate solution
\[a(t)=\sqrt{2Mt} ,\]
for the entire range of allowed values of parameter $\omega\leq
0$. It follows that radiation dominated over the unified dark
energy at early time. At late time the unified dark energy field
would start dominating over radiation, as one would expect, so
that it will be described by Eq. (16) (or any of its counterpart
solutions for $\alpha\neq 1$) which would govern the evolution of
the current universe. It appears therefore that during the time
along which the generalized Chaplygin dark-energy gas behaved like
a quintessence field at the earliest times, this was nevertheless
dominated by radiation; however, at the late times when that
generalized single gas behaves like a cosmological constant, this
dominated over radiation. We therefore hope that a more detailed
consideration of unified dark energy may lead us to solve, or at
least greatly alleviate the cosmic coincidence problem [22].
Before closing up, we mention the interesting possibility that the
unified dark energy can also behave like a quintessence field when
$a\rightarrow\infty$ and like a cosmological constant as
$a\rightarrow 0$. That would happen if $\omega <-1$, with $A$ and
$B$ taking on negative values (so keeping the current density, the
quintessence energy density and the cosmological constant all
positive), a case which is not excluded by current constraints on
the state equation for dark energy [20]. We finally note that Eq.
(2) would reduce to Eq. (1) in case that $\alpha <0$ as
$a\rightarrow 0$, so indicating that in such a case our unified
model for dark energy also included the current unified model for
dark matter in the limit of high density.

\acknowledgements

\noindent The author thanks Mariam Bouhmadi and Carmen L. Sig\"{u}enza
for useful conversations. This work was supported by MCYT under
Research Project No. BMF2002-03758.


\begin{references}
\bibitem {21} Eq. (17) can be derived from an action for
Hilbert-Einstein gravity $S=\int d^4 x\sqrt{-g}\left[(1-4\pi
G\varphi^2/3)R+L_{{\rm gas}}+L_{\varphi}\right]/16\pi G+ S$, where
$L_{{\rm gas}}$ is the Lagrangian for our generalized Chaplygin
gas, $L_{\varphi}$ is the Lagrangian for the conformally coupled
scalar field $\varphi$, and $S$ represents a boundary integral
which does not involve the generalized Chaplygin gas, specializing
to a FRW metric $ds^2=-dt^2+ a(t)^2 d\Omega_3^2$, with
$d\Omega_3^2$ the metric on the unit three-sphere.
\end{references}
\end{document}